# Fermilab Muon Campus g-2 Cryogenic Distribution Remote Control System

L. Pei, J. Theilacker, A. Klebaner, W. Soyars, R. Bossert

*Fermi National Accelerator Laboratory*
*Batavia, IL, 60510, USA*

**Abstract:** The Muon Campus (MC) is able to measure Muon g-2 with high precision and comparing its value to the theoretical prediction. The MC has four 300 KW screw compressors and four liquid helium refrigerators. The centerpiece of the Muon g-2 experiment at Fermilab is a large, 50-foot-diameter superconducting muon storage ring. This one-of-a-kind ring, made of steel, aluminum and superconducting wire, was built for the previous g-2 experiment at Brookhaven. Due to each subsystem has to be far away from each other and be placed in the distant location, therefore, Siemens Process Control System PCS7-400, Automation Direct DL205 & DL05 PLC, Synoptic and Fermilab ACNET HMI are the ideal choices as the MC g-2 cryogenic distribution real-time and on-Line remote control system.

This paper presents a method which has been successfully used by many Fermilab distribution cryogenic real-time and On-Line remote control systems.

## INTRODUCTION

Fermilab Muon g-2 experiments will examine the precession of muons that are subjected to a magnetic field. The main goal is to explore rare sub-atomic processes and make precision measurements , test the Standard Model's predictions of this value by measuring the precession rate experimentally to a precision of 0.14 parts per million. The Muon Campus experiments 1 (MC1) will house the cryogenics plant as well as the 50 foot wide Muon g-2 superconducting particle storage ring. The experiments utilize superconducting magnets that require cryogenic services. A cryogenic system to provide these services consists of four Tevatron Satellite refrigerators, four compressors, a cryogenic distribution system, and an axillary system necessary for the cryogenic system operation. One of the buildings (A0), will house the noisy vibrating equipment (four compressors) needed to operate the cryogenic plant. The other building will contain the cryogenic plants with four Tevatron Satellite refrigerator systems, Muon g-2 electromagnet ring, and an office area.

The current plan is for the MC facility to completing reassembly of all the magnet components. Soon, the coils will be cooled to near absolute zero with liquid helium so that the superconducting magnet can undergo tests and commissioning. The MC test will be used to assess the cryogenic plant performance prior to their commissioning. A layout of the entire Muon Campus facility complex is shown in FIGURE 1.

## TEST FACILITY DESCRIPTION

The cryogenic test facility is composed of A0 compressor room and MC1 helium cryogenic refrigeration, which is provided by four onsite Tevatron style of cryogenic plants. The warm compressors located at A0 Compressor Building are four MyCom brand 300KW compressors shown on the FIGURE 2. The cryogenic refrigerator plant is shown on the FIGURE 3; there are four reciprocating dry expanders and four reciprocating wet expanders. They will supply liquid helium to Muon g-2 and Mu2e solenoids. The CM1 also houses Muon g-2 solenoids ring as shown on the FIGURE 4.

Muon g-2 solenoids maximum cryogenic heat loads is liquefaction load ~ 1.4 [g/sec] and refrigeration load ~ 300 [W]. And Muon Mu2e solenoids maximum cryogenic heat loads is liquefaction load ~ 0.8 [g/sec] and refrigeration load ~ 350 [W].

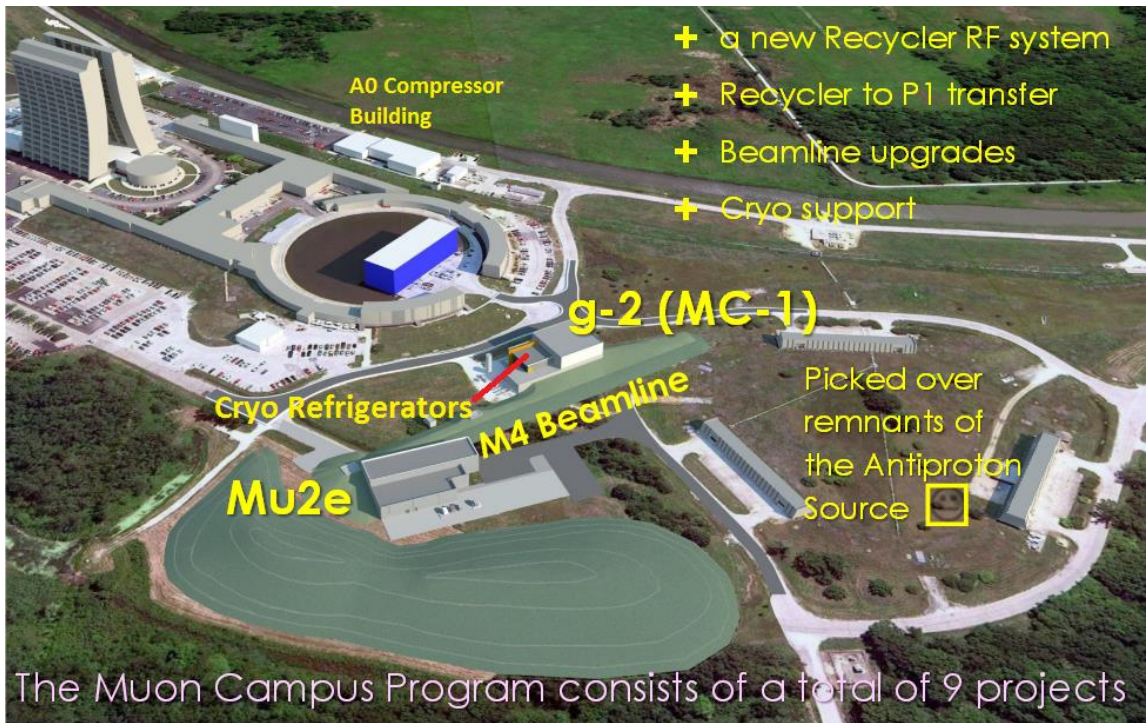
**FIGURE 1.** Layout of Muon Campus Experimenter

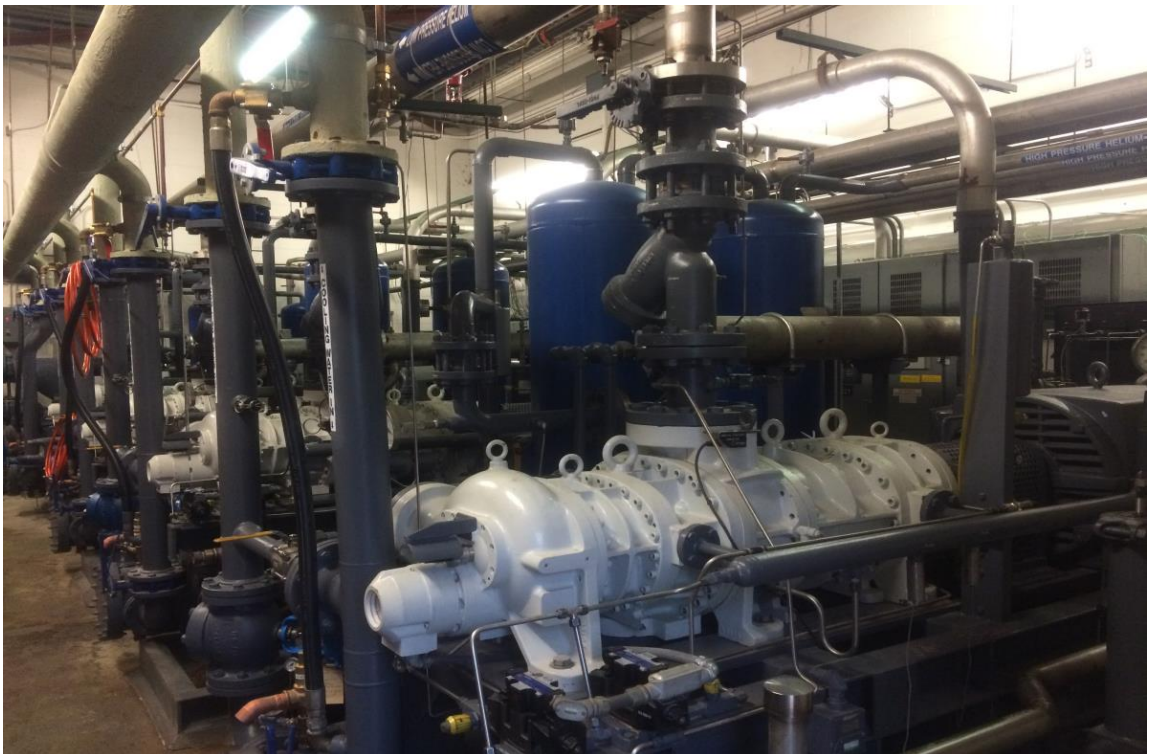
**FIGURE 2.** A0 Warm Compressor Room

Tevatron style satellite refrigerators maximum theoretical capacity is liquefaction ~ 4.2 [g/sec], refrigeration ~625 [W]. The Cryogenic System is expected to operate for 20 years with an estimated shutdown period of one month every year. The Cryogenic System shall support simultaneous steady state operation of both experiments,

Muon g-2 and Mu2e. It shall provide for independent operation of the two experiments, including transient, refrigeration, liquefaction, warm-up and cool down mode, etc.

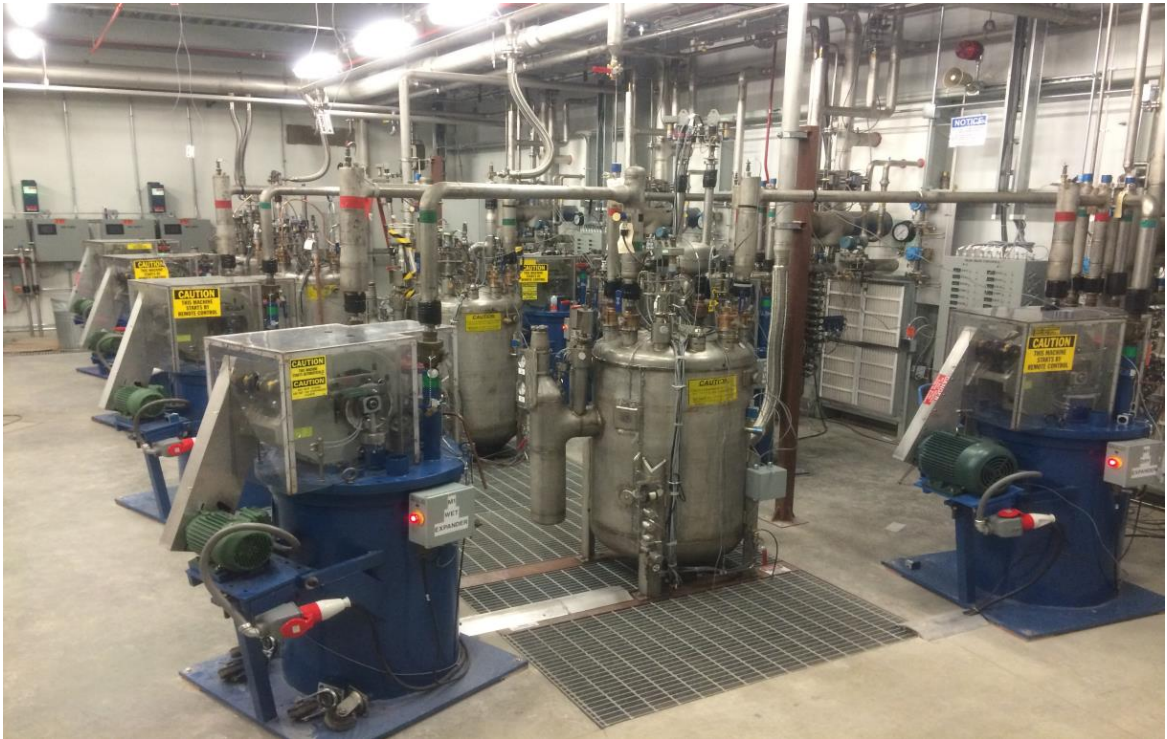
**FIGURE 3.** CM1 Four Tevatron style satellite refrigerators

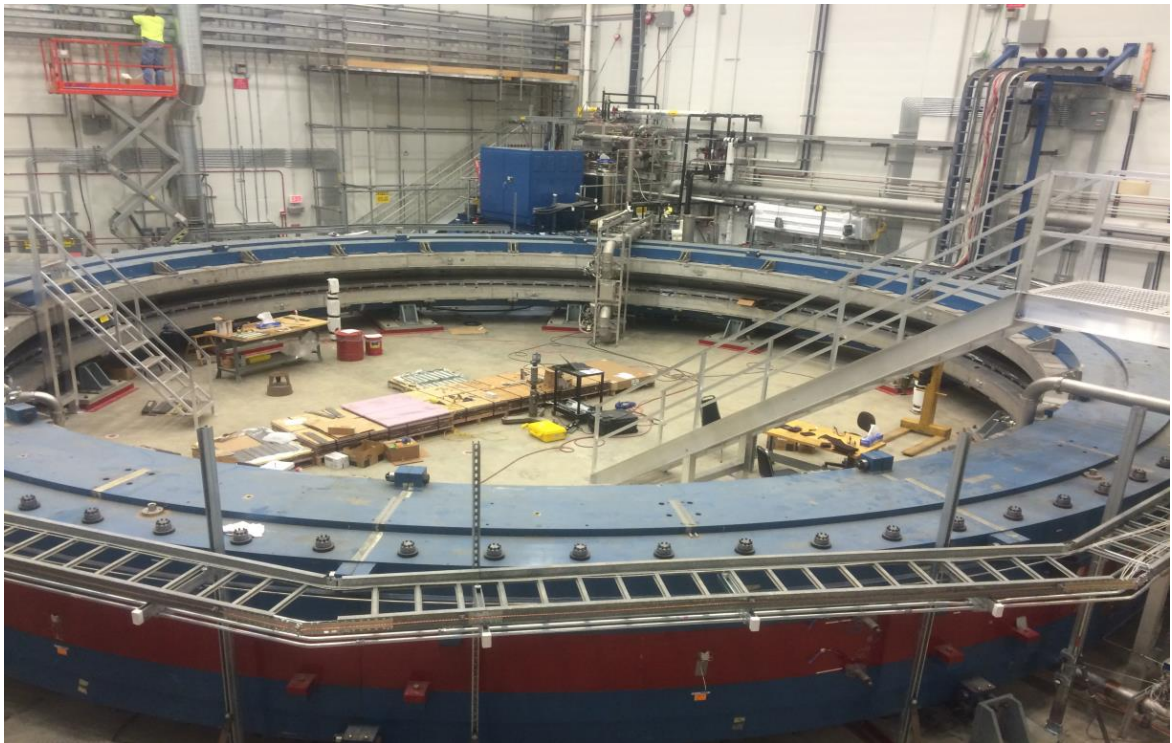
FIGURE 4. CM1 Muon g-2 solenoids ring

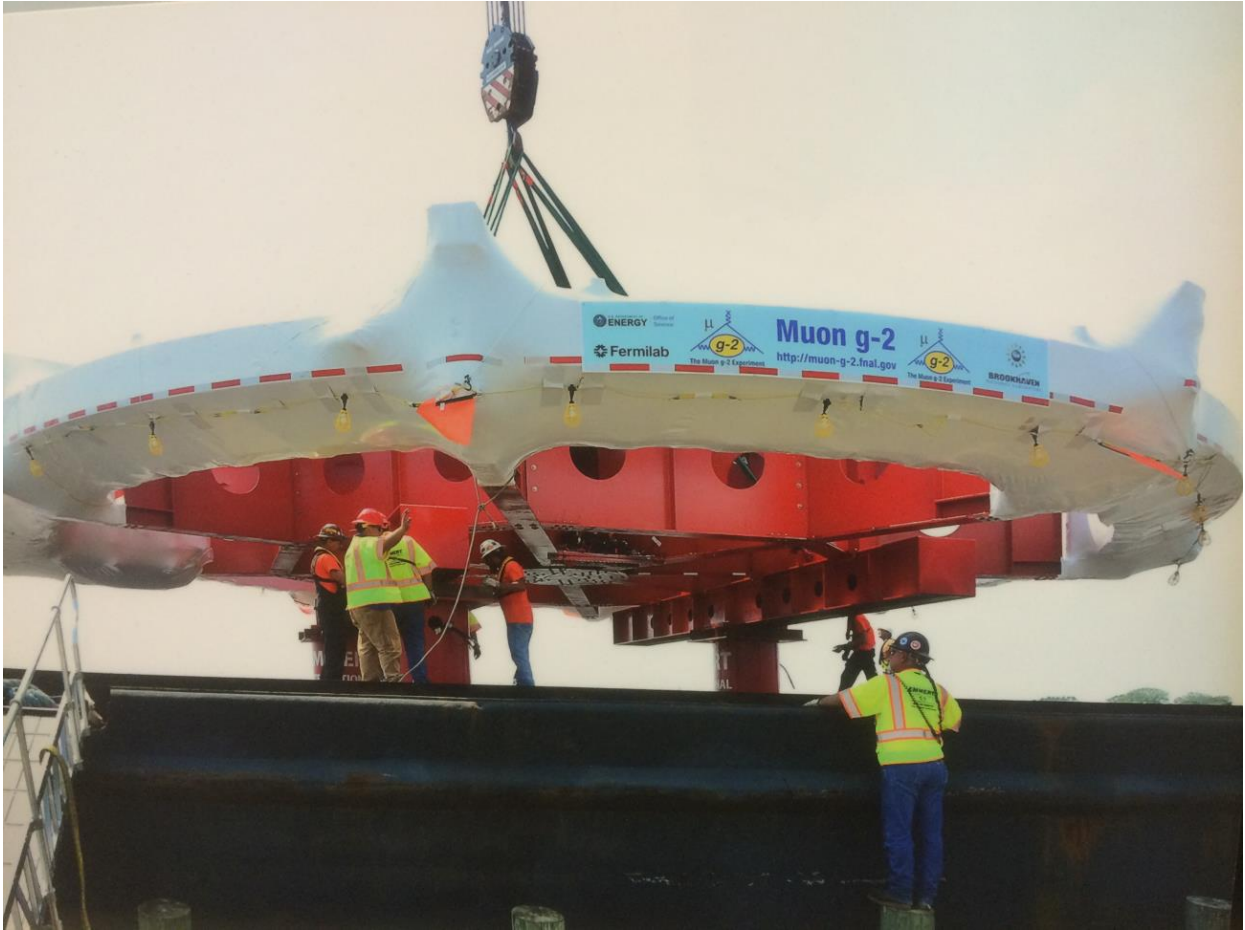
FIGURE 5. CM1 Muon g-2 solenoids ring on traveling

# DISTRUBITION REMOTE CONTROL SYSTEM

The Muon Campus 1 cryogenic system utilizes Siemens Process Control System 7 (PCS 7) controls. Its main control system consists of the Siemens System SIMATIC PCS 7 which is an integral component of Totally Integrated Automation (TIA); TIA is a unique platform for unified and customer-specific automation system. The similar control system has been successfully used for many years at Fermilab to control other cryogenic systems, e.g. CMTF. Simplified schematics for the CM1 cryogenic controls system is shown on FIGURE 6.

The multi-level distribution remote control system uses Siemens Engineering Station (ES) as its operation level; SIMATIC NET IE as its OPC (Object Linking and Embedding(OLE) for Process Control) server; Fermilab Synoptic HMI system as its Web operation and monitor level; Fermilab ACNET (Accelerator Control Network) as its archive, monitor and alarm level.

The Siemens PCS7-400 PLC acts as the central control PLC and four ET200Ms act as remote data acquisition and field I/O that are linked to the controls PCS7-400 PLC through the field control bus PROFIBUS DP. There are fifteen Automation Direct DL205/DL06 PLCs by KOYO® as its remote independent sub-control field system. Among them, six DL205s are used as local control PLCs for compressor operation, inventory management and real-time interlock protection; eight DL06s are used as local control PLC for engine operation and real-time interlock protection.

The DL205s/DL06s communicate to the PCS7-400 controls PLC through one DL205 configured as a gateway PLC. The PCS7-400 central control PLC handles with all PID LOOP control, signal conversion and logic control as well as communication between Fermilab ACNET and PCS7-400 and DL205s/DL205s. Fiber Media Converters (MC) and Scalance Ethernet Switches (SW) are used at various locations to integrate different parts of the system. The CM1 PCS7-400 central control PLC, ET200M I/O and Scalance Ethernet switch are shown on FIGURE 7.

The control of the localized equipment such as the 300 KW MyCom compressors, four reciprocating dry expanders and four reciprocating wet expanders are done using localized, self-contained, and PLC based sub-controls system which communicate directly with the PCS7 system using the fiber optic line. The localized PLCs interface with the equipment motor controller and manage the machine local interlocks. The start/stop/reset features, the remote/local control as well as a limited amount of input and output channels are also managed by this PLC Locally, a local touch panel display allows for manipulation and control of these systems and parameters.

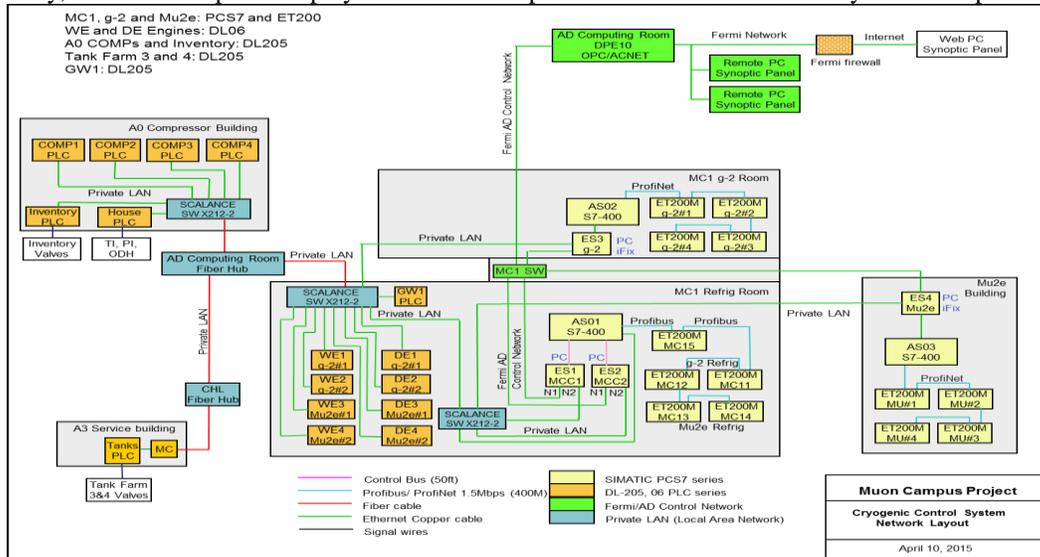

**FIGURE 6. Muon Campus Cryogenic Control System Network Layout**

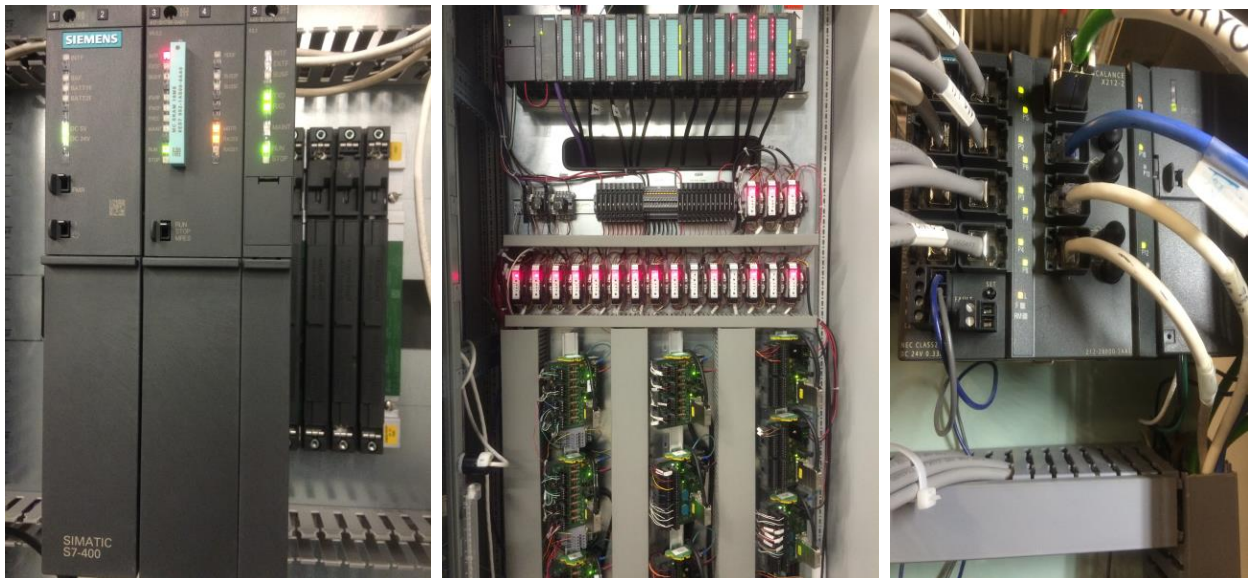

FIGURE 7. The CM1 PCS7-400 central controller, ET200M I/O and Scalance Ethernet switch

The top layer human machine interface used for the CM1 Cryogenic system is Synoptic graphic user interface HMI from Sun-Microsystem JAVA. PCS7 parameters are sent to the Fermilab ACNET through SIMATIC NET IE OPC server. The Synoptic system is a graphical interface between Fermi ACNET and the end user which uses graphical tools to display the cryogenic process. Control of the system can also be done using those tools by simply clicking on graphical components and manipulating the output. The displays are created using the Synoptic graphical builder. Synoptic also supports alarm handling and plotting packages.

This ACNET or Synoptic flexibility gives experimenters access to data from various systems in one platform for ease of plotting and data management.

The Muon Campus cryogenic control system is fully protected by safety relief valves and system configuration and in no way relies on the operation of the controls system. The controls system is used to automate the various processes through a network of control loops and logic and acts only as a secondary, supplemental safety system.

## INITIAL COMMISSIONING

Over the last half year some key components of the CM1 cryogenic system have been successfully commissioned, it including the warm compressors as well as the inventory control system. During the commissioning, four 300KW MyCom compressors are operating as show on FIGURE 8.

The inventory pneumatic control valves PVHS, PVLS and PVLB are shown on the FIGURE 8 is available to maintain pressure while compressor flow will be changed with loading valve. We used PCS 7-400 PID LOOP control to regular discharge pressure PT2 at 275 psig and suction pressure PT1 at 1.5 psig. All high sliders of four compressors are set up at 100% by Synoptic HMI manual setting while compressor running.

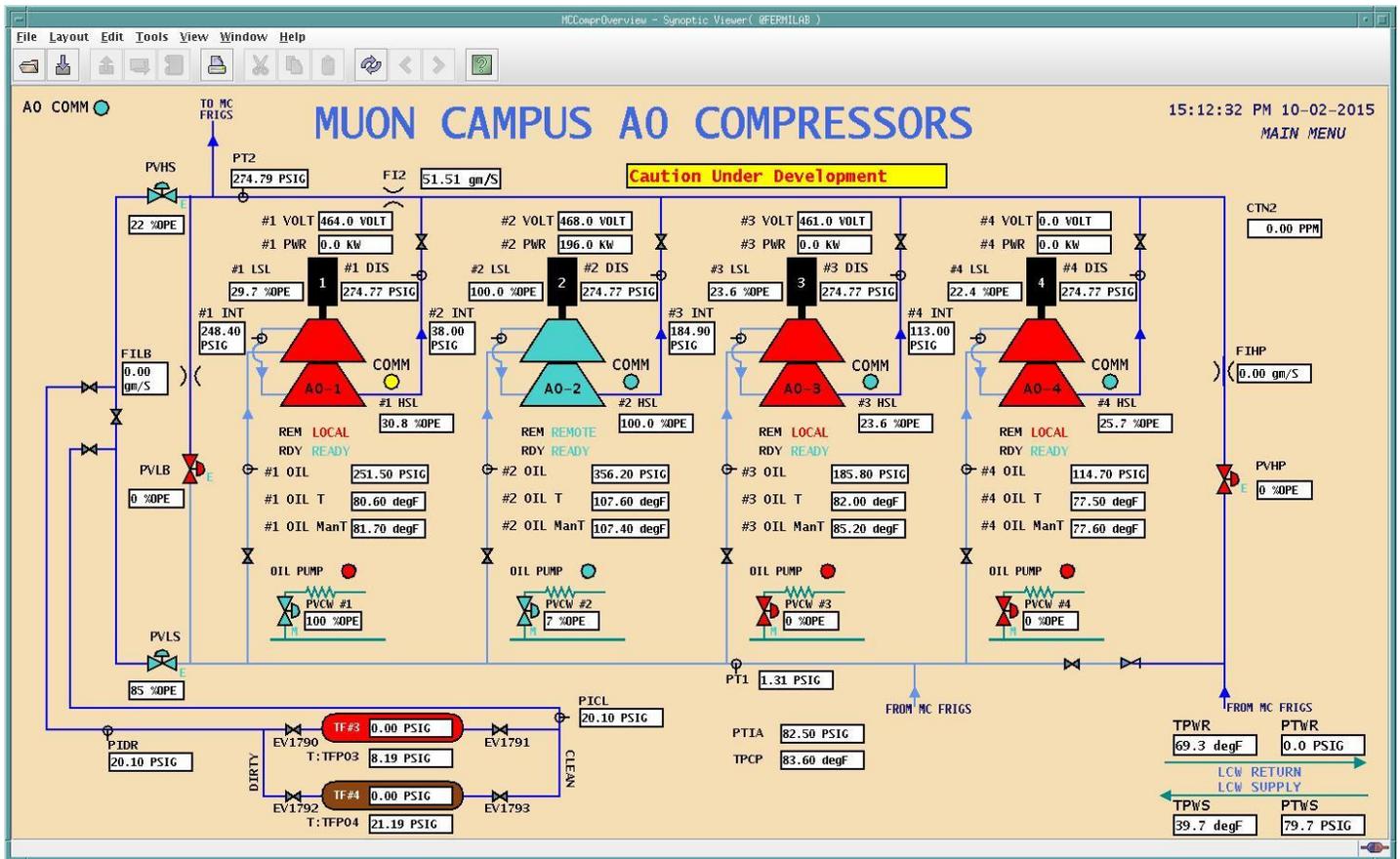

**Figure 8**. Four A0 compressors operation status is shown up by Synoptic HMI

M1 Mu2e refrigerator system is operating as show on FIGURE 9. The sustainable capacity of a single Tevatron satellite refrigerator is either: ~50 W at 4.5 K in refrigerator mode or of ~3.7 g/s of liquid helium at atmospheric pressure in liquefaction mode.

The next phase of commissioning will include cooling down the installing Muon g-2 and Mu2e in the future.

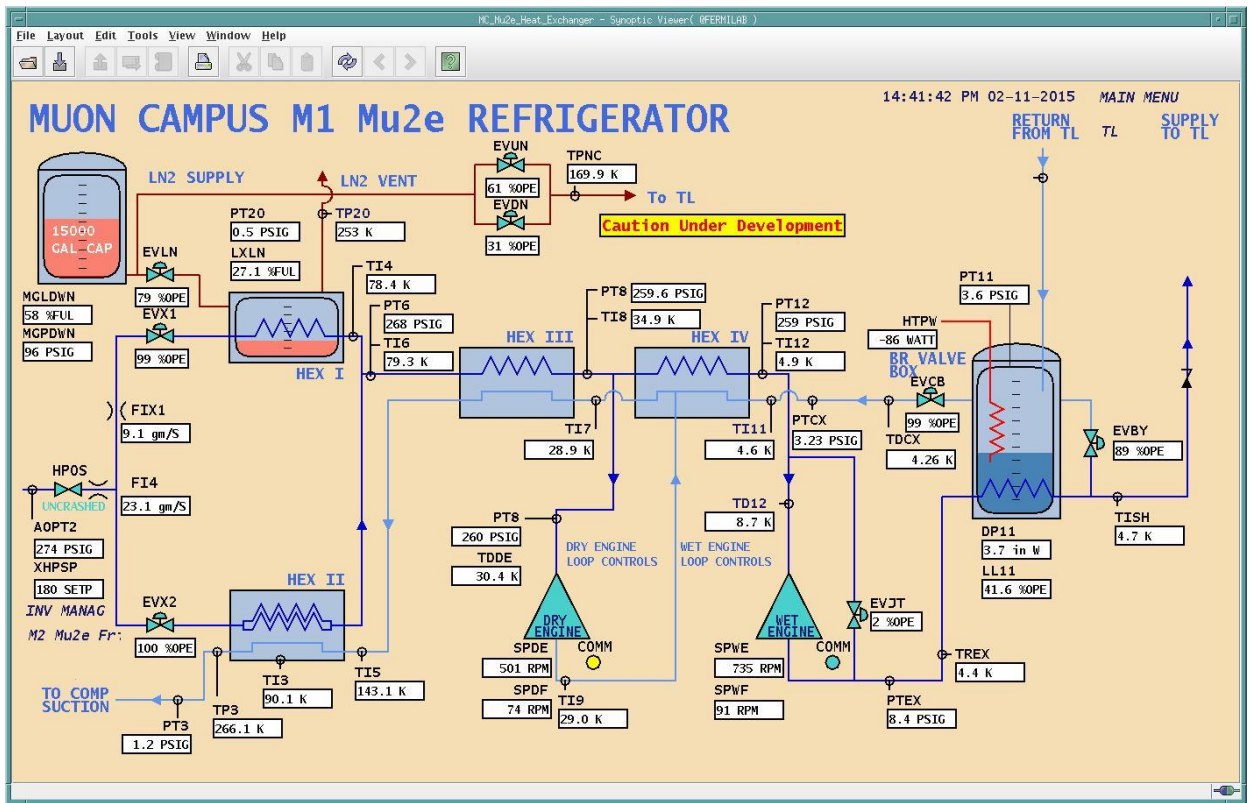

**Figure 9.** CM1 Muon g-2 refrigerator management is shown up by Synoptic HMI

## CONCLUSION

The CM1 cryogenic system is working on its schedule with the successful commissioning of several subsystems, including the Warm compressor systems, inventory control system, refrigerator system and PCS 7-400 remote distribution cryogenic control system. The commissioning of the cryogenic distribution system leading to the new Muon g-2 test will soon be underway followed by full commissioning of the Muon g-2 system. The main goal of the Muon g-2 experimenters is to test the Standard Model's predictions of this value by measuring the precession rate experimentally to a precision of 0.14 parts per million. If there is an inconsistency, it could indicate the Standard Model is incomplete and in need of revision.

## ACKNOWLEDGMENTS

This work is supported by Fermi Research Alliance, LLC under Contract No. DE-AC02-07CH11359 with the United States Department of Energy. The authors wish to recognize the dedication and skills of the Accelerator Cryogenics Department technical personnel involved in the operation of this system.